\documentclass[a4paper,10pt]{article}
\usepackage[utf8]{inputenc}
\usepackage{amsmath}
\usepackage{physics}
\usepackage{amsfonts}
\usepackage{amssymb}
\usepackage{makeidx}
\usepackage{xcolor}
\usepackage{graphicx}
\usepackage{lmodern}
\usepackage{float}
\usepackage{comment}

\newcommand{\be}{\begin{equation}}
\newcommand{\ee}{\end{equation}}

\title{Testing  Short Distance  Anisotropy in Space }
\author{Robert B. Mann$^{1,2}$, Idrus Husin$^{3,4}$,  Hrishikesh Patel$^{5}$\footnote{Correspondence to: hrishikesh.patel@alumni.ubc.ca}, \\ Mir Faizal$^{6,7,8}$, Anto Sulaksono$^3$,
Agus Suroso$^9$
\\\\
\textit{\small $^1$Department of Physics and Astronomy University of Waterloo,} \\ \textit{\small Waterloo, Ontario, N2L 3G1, Canada}
\\
\textit{\small $^2$Perimeter Institute, 31 Caroline St. N., Waterloo,} \\ \textit{\small Ontario, N2L 2Y5, Canada}
\\
\textit{\small $^3$Departemen Fisika, FMIPA, Universitas Indonesia, Depok 1624, Indonesia}
\\
\textit{\small $^4$IoT and Physics Lab, Sampoerna University, Jakarta 12780, Indonesia}
\\
\textit{\small $^5$Department of Physics and Astronomy, University of British Columbia,} \\ \textit{\small 6224  Agricultural Road,
Vancouver, V6T 1Z1, Canada}
\\
\textit{\small $^6$Department of Physics and Astronomy,
University of Lethbridge,} \\ \textit{\small Lethbridge, AB T1K 3M4, Canada}
\\
\textit{\small $^7$Irving K. Barber School of Arts and Sciences, University of British Columbia}\\
\textit{\small Okanagan Campus, Kelowna, V1V1V7, Canada}
\\
\textit{\small $^8$Canadian Quantum Research Center, 204-3002, 32 Ave,}
\\
\textit{\small Vernon, BC, V1T 2L7, Canada}
\\
\textit{\small $^9$Theoretical Physics Lab, THEPI Division, Institut Teknologi Bandung,} \\ \textit{\small Jl. Ganesha 10, Bandung, 40132, Indonesia}
}
\date{}
\begin{document}

\maketitle

\begin{abstract}
{The isotropy of space is not a logical requirement but rather is an empirical question; indeed there is suggestive} evidence that universe might be anisotropic. A plausible source of these anisotropies could be quantum gravity corrections. If these corrections happen to be between the electroweak scale and the Planck scale, then these anisotropies can have measurable consequences at short distances and their effects can be measured using ultra sensitive condensed matter systems.   We investigate how such  anisotropic quantum gravity corrections modify  low energy physics through an anisotropic deformation of the Heisenberg algebra. 
We discuss how such anisotropies might be observed using a scanning tunnelling microscope. 
\end{abstract}
\date{}

\maketitle

  The  fundamental degrees of freedom  of quantum gravity   are expected to be very different from general relativity.
 However any theory of quantum gravity, upon integrating out some  degrees of freedom  to obtain a low energy effective action, must yield general relativity. Among other things, this implies that local
 Lorentz symmetry might break due to quantum gravitational effects \cite{l1, l2}, and emerge  only as a low energy effective symmetry that is not expected   to hold at sufficiently high energies. 
Although  Lorentz symmetry is usually broken from $SO(3, 1) \to SO(3)$ \cite{mo12, mo14},  
it has been suggested that the Lorentz symmetry can also break from $SO(3, 1) \to SO(2, 1)$ due to a novel  gravitational Higgs mechanism \cite{l4, l5}. This would break the isotropy of spacetime, with potentially important measurable consequences.   
Furthermore,   quantum gravity could make spacetime discrete near the Planck scale \cite{j2, j4}, a notion   employed in loop quantum gravity \cite{lq12, lq14, lq16, lq18}.
At large scales  a continuous isotropic spacetime with  local Lorentz  symmetry is anticipated to emerge from
this discrete spacetime.   However at short distances we expect this leading order structure to be modified due to {an underlying discreteness that} is expected to break the isotropy of spacetime.  A similar phenomenon has been observed in condensed matter physics, where   isotropy  (and local Lorentz symmetry) emerges in graphene when {only the  nearest-neighbour} atom contributions are considered, whose physics can be expressed via a $(2+1)$ dimensional Dirac equation \cite{1a, 2a}.  Upon taking into account contributions from next-nearest neighbours   a deformation of the Dirac equation is observed \cite{I, Iorio:2019exr}. This deformation is consistent with the deformation produced from generalized uncertainty principle (GUP)  \cite{Kempf:1994su,p1, p2}. However, unlike the usual GUP, the GUP-like deformation produced in graphene breaks the emergent isotropy in the Dirac equation. This occurs due to the underlying discrete structure in graphene. Such breaking of  isotropy has also been observed in other condensed matter systems \cite{cond1, cond2, cond4, cond6, cond8, cond12, cond14}.

Following from this analogy, if spacetime also has a discrete structure (as has been predicted by several theories of quantum gravity), it is possible that the first order quantum corrections to the emergent continuous spacetime would also break the isotropy of space. Such corrections can be incorporated using an anisotropic GUP, where the deformation  from quantum gravity depends on the direction chosen, hence breaking the isotropy of spacetime.
Indeed, it is conceivable that observed  anisotropies   in the Cosmic Microwave Background (CMB) \cite{1z,2z} could be explained by quantum gravitational effects \cite{q12, q14} and could be produced  during inflation  \cite{1y, 2y, infl, infla}.   
Such effects would modify  field theories from their continuum limit formulations, and their leading order corrections could be expressed by an anisotropic GUP-like deformation. The possibility that an anisotropic GUP might explain observed CMB anisotropies is one of the major motivations to study the anisotropic GUP.

Spacetime anisotropy can also arise in string theory. For example,
some string-theoretic approaches to cosmology regard  the universe as a brane in a higher dimensional bulk \cite{b1, b2}, and anisotropic branes can be constructed that are dual to a deformation of super-Yang-Mills theory by a position-dependent $\theta$ term \cite{b5, b6, b7, b8}. It has also  been demonstrated that CMB anisotropies can occur in brane world models \cite{b4,b9}.  The T-duality of compact extra dimensions can be used to relate   winding modes and  Kaluza-Klein modes to such a zero point length   \cite{ze12, ze14, ze16}; this has been explicitly demonstrated  for string theory  compactified on a {torus of radius} $R$, the mass spectrum is invariant under T-duality, $R \to \alpha/ R$ and $k \to w$ (where $k$ is  the Kaluza-Klein mode and $w$ is the winding number). Thus, the information   {gained from probing length scales  below $R$ is }exactly identical to that gained above $R$;  $R$ acts as a zero point length in theory. 
The GUP can be understood as resulting from a minimal length manifest as this zero point length in spacetime \cite{Kempf:1994su,p1, p2}. 
{It is possible for $R$ to be }several orders of magnitude larger than the Planck scale (in models with large extra dimensions)  \cite{zw12, zw14}, rendering the resultant zero point length to be between the Planck and electroweak scales \cite{ze12, ze14, ze16}.   

In short,
the existence of a minimal length is a common feature in  all approaches to quantum gravity \cite{grav1, grav2}. 
Consequently,  it is possible that GUP corrections due to a minimal length greater than the Planck length will occur as a universal feature in  all approaches of quantum gravity \cite{p1, p2}. Moreover, the  minimal length in string theory as a zero point length due to T-duality,  could be related to a minimal length in discrete models of spacetime like loop quantum gravity \cite{d1, d2}.  Such a zero point length in string theory and discrete minimal length in loop quantum gravity (using polymer quantization) predict the same short distance corrections to simple low energy quantum mechanical systems \cite{po}.   

Minimal length therefore could be much greater than the Planck length in any theory of quantum gravity, leading to enhanced GUP corrections.   These enhanced GUP corrections  can be measured using ultra sensitive condensed matter systems \cite{u1,u2, u3, u4}, thereby forming a probe of anisotropic  gravitational effects. In general   GUP corrections break   Lorentz symmetry; since they are  motivated by quantum gravity,
this is not unexpected. Indeed
 Lorentz symmetry can be broken in various quantum gravitational models, based on 
loop quantum gravity \cite{lo12}, discrete spacetime \cite{lo14}, string field theory  \cite{lo16}, non-commutative geometry \cite{lo18}, and even perturbative quantum gravity \cite{lo20}. However, it is possible to constraint such Lorentz symmetry breaking using current experimental data \cite{lo25, lo26, lo22, lo24}. It may be noted that as isotropic GUP effects are usually measured using non-relativistic ultra sensitive condensed matter systems \cite{u1,u2, u3, u4}, the effects of Lorentz symmetry breaking can be neglected for such systems. The aim in this paper is to analyze the implications of an anisotropic GUP   and sketch out some possible pathways to experimentally test the presence   {of spacetime anisotropy at} short distances.  As this can again  be done using  non-relativistic ultra sensitive condensed matter systems, we can  also neglect the effects of Lorentz symmetry for anisotropic GUP.

{The standard  Heisenberg algebra $[x^i, p_j] = i\hbar \delta^i_j$ is deformed 
to incorporate minimal length in quantum gravity \cite{Kempf:1994su,p1, p2}, and can be written as
\begin{equation}
   \left[ \tilde{x}_i,\tilde{p}_j\right]=i\hbar\delta_{ij}(1+\beta\tilde{p}^2) + 2\beta \tilde p_i \tilde p_j
    \label{e1a}
\end{equation}
where $(x_i, p_j)$ are the conjugate position/momentum variable and for $\beta=0$.
The coordinate representation of the momentum operator is $p_i = - i \hbar \partial_i$ but under the deformation becomes  $\tilde p_i = -i\hbar \partial_i (1 -\hbar ^2 \beta \partial^j \partial_j )$}. Thus, we can write a map between the deformed $\tilde p_i, \tilde x^j$ and the original one as $\tilde x^j = x^j$ and $\tilde p_i = p_i (1 + \beta p^j p_j)$. 
However in this deformation we have assumed that the  deformation  is the same for all directions, and there is no fundamental reason for that assumption. 

To model anisotropic effects we therefore propose a modification of the commutation relations
\be
\left[\tilde{x}_i,\tilde{p}_j\right]=i\hbar\delta_{ij}\left(1+\beta_{k l}\tilde p_{k} \tilde p_l\right)+2i\hbar \beta_{i k} \tilde p_k \tilde p_j 
\ee
to leading order in the components of the  full deformation matrix
 $\beta_{jk}$.
For simplicity we shall henceforth assume that   off-diagonal terms vanish: $\beta_{ij} =0$ if $i \neq j$. 
Consequently we have a different 
deformation parameter  for each direction, and  by defining 
$
\beta_{xx}= \beta_{x}, $ $
\beta_{yy}=\beta_{y}, $ $
\beta_{zz}=\beta_{z} $, 
we can now  write 
the position and momentum commutation relations as
\be
\left[\tilde x_i,\tilde{p}_j\right]=i\hbar\delta_{ij}\left(1+\beta_{k}\tilde p_{k}^{2}\right)+2i\hbar\delta_{ik} \beta_k \tilde p_k \tilde p_j 
\ee  
which results in different minimal lengths in each direction
\begin{eqnarray}
\left(\Delta x\right)_{\text{min}}&=&\hbar\sqrt{\beta_x}=\sqrt{l_P\hbar \beta_{0x}}\nonumber \\
\left(\Delta y\right)_{\text{min}}&=&\hbar\sqrt{\beta_y}=\sqrt{l_P\hbar \beta_{0y}} \\
\left(\Delta z\right)_{\text{min}}&=&\hbar\sqrt{\beta_z}=\sqrt{l_P\hbar \beta_{0z}},\nonumber 
\end{eqnarray}
where $\beta_{i}={\beta_{0i}l_{P}}/{\hbar}$. The resulting parameter set $(\beta_{0x},\beta_{0y},\beta_{0z})$  
describes the  anisotropic GUP.  
The anisotropic deformation of the momentum operator is 
\begin{eqnarray}\label{e4}
\tilde{p}_i=\left(1+ p^j\beta_{jk}p^{k}\right)p_{i}=\left(1+\beta_xp_{x}^{2}+\beta_yp_{y}^{2}+\beta_zp_{z}^{2}\right)p_{i}
\end{eqnarray}


Now  using (\ref{e4}),  the Hamiltonian now can be written
\begin{eqnarray}\label{Ham}
H&=&\frac{\tilde p^2}{2m}+V(\vec{r}) \nonumber \\
&\approx& \left(\frac{p^{2}}{2m}+V\left(\vec{r}\right)\right)+\frac{\beta_k}{m}p_{k}^{2}p^2 
\end{eqnarray}
to first order in the correction term.  
Although this correction term was motivated from quantum gravity considerations, it  universally corrects all low energy quantum mechanical systems.   
 The Hamiltonian  \eqref{Ham} for the anisotropic GUP can be written as
\begin{eqnarray}\label{eH}
H&=&-\frac{\hbar^2}{2m}\nabla^2+V+\frac{\hbar^4}{m}\beta_{k}\partial_{k}^{2}\nabla^2\nonumber\\
&=&\left(-\frac{\hbar^2}{2m}\nabla^2+V\right)+\frac{\hbar^4}{m}\nabla^2\tilde{\nabla}^2\nonumber\\
&=&H_{0}+H_{p}
\end{eqnarray}
where we have defined the anisotropic Laplace operator 
\begin{eqnarray}
\tilde{\nabla}^2=\beta_{x}\partial_{x}^{2}+\beta_{y}\partial_{y}^{2}+\beta_{z}\partial_{z}^{2}
\end{eqnarray}
and 
$H_{0}=-\frac{\hbar^2}{2m}\nabla^2+V$. 
Now to understand the effects  of such a deformation on the behavior of quantum systems, we need to first analyze its effects on the continuity equation.  
The probability density and current are 
\begin{eqnarray}\label{un}
\rho = \Psi \Psi^{*} \qquad
\vec{J}_{0}=\frac{i\hbar}{2m}\left(\Psi\vec{\nabla}\Psi^{*}-\Psi^{*}\vec{\nabla}\Psi\right),
\end{eqnarray}
and using  the Schr\"{o}dinger equation
$H\Psi=i\hbar\partial_{t}\Psi$ 
we obtain
\begin{eqnarray}\label{conS}
\partial_{t}\rho+\vec{\nabla}.\vec{J}&=&
\partial_{t}\rho+\vec{\nabla}.\left(\vec{J}_{0}+\vec{J}_{p}\right)\nonumber \\
&=&\frac{i\hbar^3}{m}\left[\tilde{\nabla}^2\Psi^{*}\nabla^{2}\Psi-\tilde{\nabla}^2\Psi\nabla^{2}\Psi^{*}\right]. 
\end{eqnarray} 
where the additional term in the modified non-local  probability current is 
\begin{eqnarray}\label{per}
\vec{J}_{p}=\frac{i\hbar^3}{m}\left[\Psi^{*}\vec{\nabla}\left(\tilde{\nabla}^2\Psi\right)-\Psi\vec{\nabla}\left(\tilde{\nabla}^2\Psi^{*}\right)+\left(\tilde{\nabla}^2\Psi^{*}\right)\vec{\nabla}\Psi-\left(\tilde{\nabla}^2\Psi\right)\vec{\nabla}\Psi^{*}\right].
\end{eqnarray}
$$
$$

We observe the rather stiking result that the anisotropic GUP violates conservation of probability current, and hence particle number. Although an anistropic GUP is expected from an underlying anistropic discreteness of spacetime due to quantum gravity,  this situation is quite unlike that of local models on anistropic lattices. It is due to the intrinsic non-locality of the anisotropic GUP, and has been observed in other situations where models with non-local terms, such non-local motion of the particles violate the 
local non-conservation of probability current  \cite{fr14, fr16, fr18, fr20, fr22}. For the anisotropic GUP we are considering, this violation will not occur if $\beta_x=\beta_y=\beta_z$ (i.e. isotropy is restored), if the wavefunction is either pure real or pure imaginary, or if its Laplacian vanishes. However in generic situations it does occur.

We can investigate the   global conservation of probability by defining 
\begin{equation}
Q =   \int \rho \, dv  
\end{equation}
and  writing  
\begin{equation}
\frac{d Q}{dt} = \frac{\partial }{\partial t} \int \rho \, dv = -
 \int dS \cdot \left( \vec{J}_0 + \vec{J}_p \right) + 
 \frac{i\hbar^3}{m} \int \left[\tilde{\nabla}^2\Psi^{*}\nabla^{2}\Psi-\tilde{\nabla}^2\Psi\nabla^{2}\Psi^{*}\right] dv
\end{equation}
Here $Q$ is only conserved  if the total flux across the surface due to the local and non-local parts of the probability current is cancelled by the volume term. If the 
falloff of the current terms is sufficiently rapid, then the flux term will vanish and particles will be generated from the volume term.

We expect that this is a generic quantum gravity effect, if quantum gravity does indeed induce an anisotropic GUP. A fully self-consistent quantum theory of gravity will presumably include additional terms that  will yield particle creation/annihilation effects due to such anisotropic effects.  Lacking any such theory at present, the anisotropic GUP indicates that quantum gravity effects lead to  very small
(anisotropic) violations of quantum mechanical probability. However, this situation is not without precedent. Other examples of non-local models with  local non-conservation of probability current  are fractional Schrödinger equation \cite{fr10, fr12}, certain wave packets in a harmonic potential \cite{10fr}, fractional Feynman-Kac equation for non-Brownian functionals \cite{16fr}, Levy flights in non-homogeneous media \cite{18fr}, vicious Levy flights \cite{24fr}, subrecoil laser cooling \cite{22fr}, hydrodynamic superdiffusion  in graphene \cite{12fr}, coupled non-linear Schrödinger equations \cite{20fr},  and  certain resonant modes \cite{14fr}.

The full empirical implications of non-conservation of probability current for the anisotropic GUP remain an interesting subject for future investigation.  Here we consider one such implication, namely that local anisotropic non-conservation of probability current causes an anisotropic non-local motion of the particle. Since such non-local anisotropic corrections can occur universally in low energy quantum mechanical systems, we will investigate this issue of non-conservation and its practical implications using a concrete example. We consider in particular the motion of a particle (tunneling) through a potential barrier in a scanning tunneling microscope (STM) experiment. We expect that the anisotropy would render the transmission coefficient to be direction dependent and such directional behavior could then be experimentally observed.

To calculate the anisotropic GUP corrections consider the potential barrier  
\begin{eqnarray}
V=
\begin{cases}
V_0 & 0 \leq \tilde{x}\leq a \\\\
0 & \mathrm{otherwise}
\end{cases}
\end{eqnarray}
where 
\begin{eqnarray}
\tilde{x}=x\cos\theta+y\sin\theta, &&
\tilde{y}=-x\sin\theta+y\cos\theta
\end{eqnarray}
with $\theta$ parametrizing the angle of the barrier relative to the preferred $x$-axis.

If the particle is moving in the $\tilde{x}$ direction the wave function is given by $\Psi=\Psi\left(x'\right)$, and   we obtain
\begin{eqnarray}
-\frac{\hbar^2}{2m}\nabla^2\Psi+V\Psi+\frac{\hbar^4}{m}\beta_{k}\partial_{k}^{2}\nabla^2\Psi&=&-\frac{\hbar^2}{2m}\frac{\partial^2\Psi}{\partial {\tilde{x}}^2}+V\Psi+\frac{\hbar^4}{m}\beta_{1}\partial^{4}_{\tilde{x}}\Psi = E\Psi
\nonumber \\
\end{eqnarray}
for the one dimensional anisotropic Schr\"{o}dinger equation,
with $\partial^{4}_{\tilde{x}}={\partial}/{\partial \tilde{x}^4}$, and $
\beta_1=\beta_{1}\left(\theta\right)=\beta_x\cos^2\theta+\beta_y\sin^2\theta$ -- 
the GUP parameter is now a function of the angle  $\theta$.
Solving the equations in each potential region above, we obtain the tunneling coefficient
\begin{eqnarray}\label{tunco}
T=\frac{1}{1+\frac{\left(\tilde{k}_{1}^{2}+\tilde{k}_{2}^{2}\right)^2\sinh^{2}\left(\tilde{k_{2}}a\right)}{\left(2\tilde{k}_{1}\tilde{k}_{2}\right)^2}}
\end{eqnarray}
with anisotropic GUP corrected wave-number  $\tilde{k}_{1}$ and $\tilde{k}_{2}$
\begin{eqnarray}\label{y1}
\tilde{k}_{1}=k_1\left(1-\beta_{01}l_{P}^{2}k_{1}^{2}\right),
&& 
\tilde{k}_{2}=k_2\left(1+\beta_{01}l_{P}^{2}k_{2}^{2}\right),
\end{eqnarray}
where $k_{1}$ and $k_2$ are the usual wave-number 
\begin{eqnarray}
k_{1}=\sqrt{\frac{2mE}{\hbar^2}},
&&
k_{2}=\sqrt{\frac{2m\left(V_0-E\right)}{\hbar^2}}.
\end{eqnarray}

 The interpretation of Eq. \eqref{tunco} is a bit subtle. Consider an experiment emitting particles toward a barrier, with a detector on the other side of the barrier.  If the GUP were isotropic, there would be no change in the transmission coefficient \eqref{tunco} as the entire experiment is rotated through $2\pi$. By contrast, the anisotropic GUP predicts that the transmission coefficient will change as the experiment is rotated about the $z$-axis, violating local Lorentz invariance.

A scanning tunneling microscope (STM) could be an ideal system for measuring (or constraining)
this effect.  If we consider anisotropic GUP corrections to the STM experiment, then we would expect that the tunneling (transmission) probabilities 
differ as the experiment is rotated.   These differences in probabilities  depend on several parameters ( like $\beta_0, k_1$ and $k_2$), and so  we need to make some assumptions that will simplify our calculations while still adhering to most practical aspects of such a system.

If anisotropy exists ($\beta_{0x} \neq \beta_{0y}$) then, without loss of generality, we can assume, $\beta_{0x}< \beta_{0y}$, which in turn would imply $\beta_{0x}\leq \beta_{01} \leq \beta_{0y}$. Furthermore, we can also assume for simplicity that $k_1 = k_2$, which is physically feasible. Under these assumptions, we get  
\begin{equation}
T = \frac{1}{1 + Z \sinh^2(\tilde{k}_2a)}
    \label{trans_sim}
\end{equation}
where 
\begin{equation}
    Z = \left[\frac{1 + \beta_{01}^2 l_p^4 k_2^4 }{1 - \beta_{01}^2 l_p^4 k_2^4}\right]^2
    \label{Z}
\end{equation}
parameterizes the effect of the anisotropic GUP.

At present the value of $\beta_0$ is constrained to be about $\beta_{0i}  < 10^{21}$ \cite{p1}. Given this bound, the constraints on the experimental parameters is rather extreme in the case of tunnelling of electrons (where $k_1=k_2$).  Writing $\epsilon = \beta_{01}^2l_p^2k_2^4$ for $\epsilon<<1$, we have $Z =( ({1+\epsilon})/({1-\epsilon}))^2 \simeq(1+4\epsilon)$. In order to be able to empirically probe such effects, we must have 
$k_2 \gtrsim {\epsilon^{1/4}}/{\sqrt{\beta_0} l_P}\approx 10^{23} \epsilon^{1/4}$ cm$^{-1}$, implying extremely high energies are necessary.
Furthermore, the potential well will have to be extremely narrow $a \approx 10^{-22}$ in order for the
$\sinh^2(\tilde{k}_2a)$ term to not fully suppress $T$. We illustrate in Fig. 1 how the transmission coefficient varies as a function of angle for parameter choices in this range.
 
 Such extreme parameter choices make the feasibility of any experiments enormously challenging.  We can ameliorate  such extremities
by going to another limit, with $k_1>>k_2$. This brings the transmission coefficient $T$ close to 1, making measurement of anisotropy more feasible.
A computational algorithm can be used to find the optimal set of parameters in the high dimensional parameter space for observing the effect.

\begin{figure}[h!]
            \begin{center}
            $%
            \begin{array}{cccc}
            \includegraphics[width=110 mm]{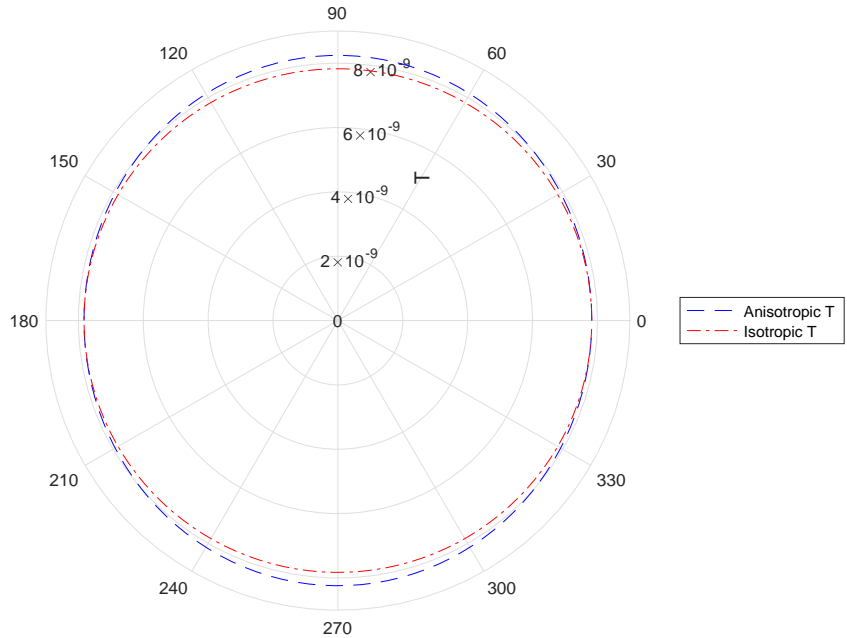}
            \end{array}%
            $%
            \caption{Variation of Transmission coefficient with angle due to anisotropic corrections. Parameter: $k_1=k_2=10^{23}$, $a=10^{-22}$, $\beta_{0x}=10^{21}$, $\beta_{0y}=10^{19}$ }
            \end{center}
\label{fig}            
\end{figure}

In this letter, we have proposed an anisotropic GUP, which  breaks the isotropy of space at short distances. We have   observed that this   anisotropic GUP  causes an effective non-local motion of quantum particles, and which  in turn causes a local non-conservation of probability current. As this deformation was proposed to occur due to low energy consequences of quantum gravitational effects, it   affects all quantum mechanical systems.  We have proposed that it can be detected using ultra precise measurements of quantum mechanical systems. In fact, we  have explicitly  proposed  that STM can be used as such a system to detect this anisotropic GUP.


We close by commenting on the implications of our results for Lorentz covariance. In the isotropic GUP there is an intrinsic minimal length without a minimal time,  breaking   spacetime covariance  \cite{Kempf:1994su,p1, p2}. Such    breaking   has been constrained from present observations \cite{lorentza, lorentzb}.  It may be noted that  such effects are not important as GUP deformation is  usually studied for  high precision and low energy non-relativistic quantum mechanical systems  \cite{u1,u2, u3, u4}. 
However  covariant formulations of the GUP exist that  contain an intrinsic minimal time, and this does not break   Lorentz symmetry \cite{cova1, cova2}. 

However  unlike the isotropic GUP, it is not possible to incorporate  additional  structure in the anisotropic GUP to restore Lorentz symmetry. This means that Lorentz-symmetry breaking is a generic prediction of the anisotropic GUP, and must be either determined or constrained from experiment, similar to what is done in
DSR  \cite{dsr1, dsr2} and Horava–Lifshitz gravity \cite{lifhitz1, lifshitz2}.  Investigating such constraints for the anisotropic GUP would be  interesting   as there is an  abundance of relevant experiments, including  gravitational waves   \cite{lorentz1}, ultrahigh-energy cosmic rays \cite{lorentz2},   lunar laser ranging \cite{lorentz4}, 
frequency differences between  Zeeman masers \cite{lorentz6}, and radio-frequency spectroscopy of atomic dysprosium  \cite{lorentz8}.   

One interesting avenue of study is an analysis of  the  cosmological and astrophysical implications of the anisotropic GUP. For example,   CMB anisotropies \cite{1z,2z}   could be due either to anisotropies in the electromagnetic field or  gravitational waves or both. It is possible to obtain corrections to Maxwell's equations from GUP, by requiring the GUP deformed matter fields to be invariant under $U(1)$ gauge symmetry \cite{k1}. This approach can also be extended to non-abelian gauge theories \cite{k2},
and even other fields like gravity (as it can be considered as a gauge theory of Lorentz group) \cite{k1}. Furthermore, it has been demonstrated that this formalism can be used to obtain corrections to these fields under other deformations of the Heisenberg algebra \cite{k5, k6}.  A similar program could be carried out for the anisotropic GUP to see what its experimental implications are. 

\section*{Acknowledgements}
R.B.M. was supported in part by the Natural Sciences and Engineering Research Council of Canada and by AOARD Grant FA2386-19-1-4077. I.H. and Ag.S. were supported in part by Riset ITB 2020 and PDUPT DIKTI 2020.

\end{document}